\documentclass{svjour3}                     % onecolumn (standard format)
\usepackage{isolatin1}
\usepackage{graphicx}

\smartqed  % flush right qed marks, e.g. at end of proof

\newcommand{\un}{~\mathrm}
\newcommand{\ie}{{\em i.e. }}
\newcommand{\eg}{{\em e.g. }}
\newcommand{\unm}{~\mu\mathrm{m}}

\journalname{International Journal of Fracture}

\begin{document}

\title{Crack opening profile in DCDC specimen}

\author{Gaël Pallares \and Laurent Ponson \and Antoine Grimaldi \and Matthieu George \and Guillaume Prevot \and and Matteo Ciccotti}

\institute{G. Pallares \and A. Grimaldi \and M. George \and G. Prevot \and M. Ciccotti
              \at Laboratoire des Colloïdes, Verres et Nanomatériaux, CNRS, Université Montpellier 2, France \\
              Tel.: +33-4-67143529\\
              Fax: +33-4-67144637\\
              \email{ciccotti@lcvn.univ-montp2.fr}
              \\
              \\
              \emph{Present address:} of G. Pallares \\
              CEA, IRAMIS, SPCSI, Grp. Complex Systems \& Fracture, F-91191 Gif Sur Yvette, France  \\ %  if needed
           \and
           L. Ponson \at Division of Engineering and Applied Science, California Institute of Technology, Pasadena, CA 91125, USA \\
              \email{ponson@caltech.edu}           %  \\
}

\date{Received: date / Accepted: date}

\maketitle

\begin{abstract}
The opening profile of the cracks produced in the Double Cleavage Drilled Compression (DCDC) specimens for brittle materials is investigated.
The study is achieved by combining Finite Element simulations of a DCDC linear elastic medium with experimental measurements by crack opening interferometry on pure silica glass samples.
We show that the shape of the crack can be described by a simple expression as a function of the geometrical parameters of the sample and the external loading conditions. This result can be used to measure accurately in real time relevant quantities during DCDC experiments, such as the crack length or the stress intensity factor applied to the specimen.
\end{abstract}

\section{Introduction}
\label{intro}

The Double Cleavage Drilled Compression (DCDC) sample refers to a
parallelepipedic column with a circular hole drilled through its
center that is subjected to axial compression. This geometry
provides many advantages in studying the fracture of brittle
materials. Janssen originally introduced such a fracture test for
measuring the fracture toughness of glass (Janssen 1974). Under a
uniform axial compression, the Poisson effect produces a tensile
stress concentrated around the central hole. This induces the
initiation of two symmetric mode I cracks generated at each crown
of the hole and propagating along the mid-plane of the sample as
the axial compression is increased (see Figs. \ref{Fig1} and
\ref{fig:Fig7}). The high stability of the crack propagation has
made of the DCDC a rather popular fracture test geometry used in
various contexts including studies on sub-critical crack
propagation in oxide glasses (Célarié et al. 2003; Bonamy et al.
2006; Grimaldi et al. 2008; Fett et al. 2008), on crack healing in
polymers (Plaisted et al. 2006) or on failure behavior under mixed
mode loading for linear elastic materials (Larnder et al. 2001;
Fett el al. 2005).

The present study presents a coupled theoretical and experimental
investigation of the crack opening profile in the DCDC specimen. A
2D Finite Element (FE) simulation based on linear elasticity is compared with a direct
measurement by crack opening interferometry on a silica glass specimen.
The results are expressed in a simple form as a function of the geometrical
parameters of the specimen following a Williams expansion series.
This experimental technique, combined with the results of the
simulations, allows the direct measurement of the stress
intensity factor driving the crack propagation in the DCDC specimen.
%\textbf{in the case of brittle elastic material for which can be describrd by the linear elastic fracture mechanics}.
In addition, it provides a solid base for the interpretation of
recent experiments on the nanoscale capillary condensation of
water inside sharp cracks in glass (Grimaldi et al. 2008).
We should stress that this linear elastic modelling only applies for very brittle materials like glass (for which the DCDC sample is mostly adapted). When dealing with more compliant materials such as PMMA the second order terms due to rotation become very relevant and should be considered by a nonlinear modelling (Plaisted et al. 2006).

\section{Finite Element simulation of DCDC specimens}

\subsection{Loading configuration and finite element mesh}

\begin{figure}[!h]
\begin{center}
\includegraphics[width=8.4cm]{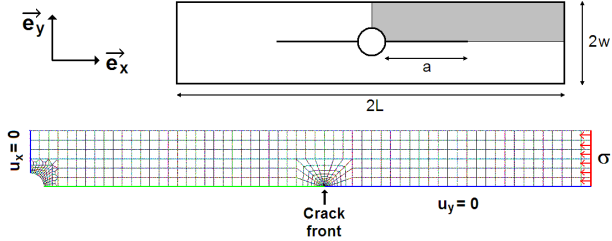}
\caption{Sketch of the DCDC specimen and details of the numerical
simulation: The mesh used in the finite element calculations is
only one quarter of the whole sample represented in grey on the
scheme of the geometry. The element size decreases exponentially
when approaching the crack tip, so that deformations of the crack
profile down to one nanometer from the crack tip can be computed}
\label{Fig1}
\end{center}
\end{figure}

The finite element simulation used to estimate the crack opening profile of the DCDC specimens is based on a 2D plane strain
analysis implemented by the finite element code CAST3M\footnote{CAST3M software is developed by CEA-Saclay,
France. Reference web page: http://www-cast3m.cea.fr/cast3m/index.jsp}.
The mesh is constituted of linear elastic isotropic four nodes elements.
The plane strain condition is chosen to model the behavior of the sample in its mid-plane, where the experimental measurements are made.
The details of the sample loading configuration and the mesh geometry are shown in Fig. \ref{Fig1}. The sample consists of a
prism of dimensions $2w \times 2t \times 2L$ with a cylindrical cross hole of radius $R$ drilled through the specimen (thickness
$2t$). The sample is loaded with a compressive stress $\sigma$ applied to the two opposite faces. The Poisson's ratio and the
Young's modulus were set to $\nu = 0.17$ and $E = 72\un{GPa}$ respectively to represent the properties of silica glass used for
the experiments. However, the influence of $E$ and $\nu$ on the simulated displacement fields and the stress intensity factor $K_I$ are explicitly given in the following. There are effects of $\nu$ on the computed value of $K_I$ that are not captured by the analytical description proposed, but they remain inferior to 1\% in the range $0.1<\nu<0.4$.
During the test, two symmetric cracks propagate from the central hole in opposite directions along the midplane of the sample. The difference in the lengths of the two cracks is less than 1\% and we will call $a$ the average crack length as measured from the side of the hole.
Thanks to symmetry of the geometry, one can limit the simulation to one quarter of the specimen (composed of about 3500 nodes), imposing specific boundary conditions to the system (cf. Fig. \ref{Fig1} bottom), i.e. $u_x = 0$ along the symmetry plan normal to $\vec{e_x}$ and $u_y = 0$ along the one normal to $\vec{e_y}$ on the uncracked ligament.
%No displacement constraint is imposed on the stress application faces, meaning that friction at the platen/sample surfaces is neglected. This effect might only affect very long cracks that might be out of the domain retained here, as confirmed by the rapid acceleration of the crack close to the sample side.
No displacement constraint is imposed on the stress application faces, meaning that friction at the platen/sample surfaces is neglected. The presence of frictional confinement might partially suppress the $K_I$ rise for cracks approaching the end of the sample (cf.\ Fig. \ref{Fig2} a), but the rapid acceleration of the crack at the end of the tests is indeed experimentally observed. In any case, the friction effect only affects very long cracks that are out of the domain retained here.
The gap between the millimeter length scale of the external mechanical loading and the nanometric length scale of the crack opening near the tip requires an adapted mesh for the calculation of stress and displacement fields.
To analyze such a multi-scale problem without exceeding reasonable computational times, the elements of the mesh are chosen so that their size decreases exponentially while approaching the crack tip, down to a minimum element size of 0.01 nm. This allows reproducing correctly the
asymptotic square root crack opening displacement profile predicted by linear elasticity down to a nanometer distance from the crack tip.

To investigate the influence of the geometrical parameters on the crack opening profile, a first series of meshes was designed with
seven values for the hole radius $R$ ranging from $0.4\un{mm}$ ($w/R = 5$) to $0.8\un{mm}$ ($w/R = 2.5$), keeping the width and
length constant with $w=2\un{mm}$ and $L=20\un{mm}$. The effect of the specimen width on the opening profile is studied using another
series of seven meshes, varying $w$ from $1\un{mm}$ to $4\un{mm}$ while keeping radius and length constant with $R=0.8\un{mm}$ and
$L=20\un{mm}$. For each new geometry, the crack length $a$ was changed between 2.5 and $18.0\un{mm}$ with $0.1\un{mm}$ steps.

\subsection{Evaluation of the stress intensity factor}

\begin{figure}[!h]
\begin{center}
\includegraphics[width=7.8 cm]{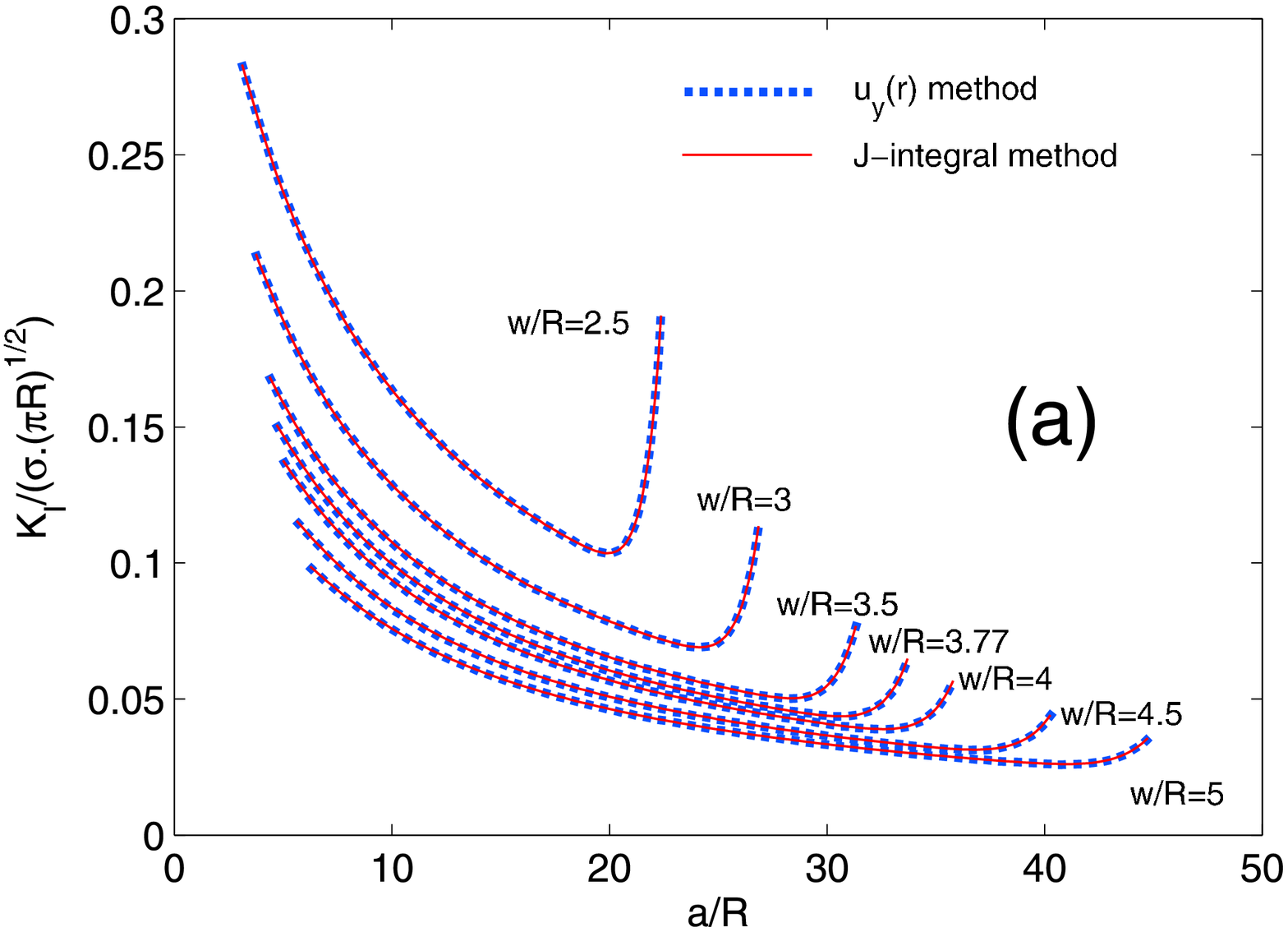}
\includegraphics[width=8.4 cm]{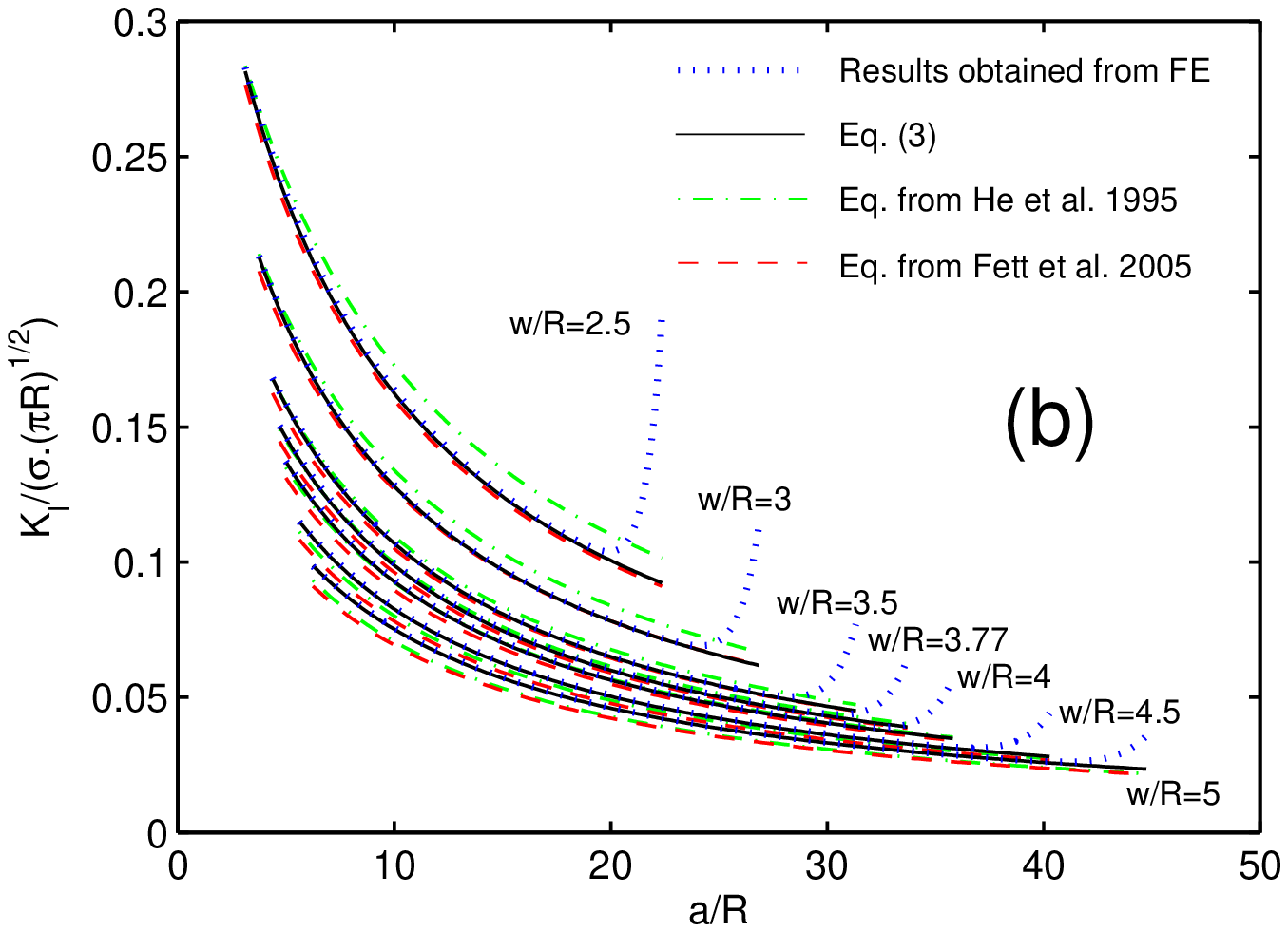}
\caption{(a) Comparison of the values of $K_I$ obtained by FE simulation from the crack opening profile (blue) using Eq. (\ref{Eq1}) and the J-integral method (red); (b) Comparison of the values of $K_I$ obtained in our simulations (blue) with the formula proposed here in Eq. \ref{eq:KInew} (black) and the ones proposed in He et al. (1995) (green) and Fett et al. (2005) (red)}
\label{Fig2}
\end{center}
\end{figure}

Once the stress, strain and displacement fields are calculated for the mesh corresponding to a given crack length $a$, the elastic
displacement $u_y(X) = u_y(x=-X,y=0)$ ($x<0$) along the crack lip is measured as a function of the distance $X$ from the crack tip. Let us note that
with these notations, the full opening between the two crack lips
is given by $2\,u_y$. The stress intensity factor $K_I$ applied to
the specimen is calculated using two different methods. At first,
the crack opening profile is fitted using the Irwin equation
(Ewalds 1985)
\begin{equation}\label{eq:IrwinEquation}
u_y(X)= {K_I \over E'} \sqrt{ 8X \over \pi}
\label{Eq1}
\end{equation}
with $E' = E / (1 - \nu^2)$ due to plane strain condition. The regression is done in the very vicinity of the crack tip, typically in the range $1\un{nm}< X <1\unm$ where the crack profile follows a square root behavior as shown in the following. On the other hand, the J-integral method (Rice 1968) is computed, providing the mechanical energy release rate $G_I$ related to the stress intensity factor by the expression $G_I=K_I^2/E'$ (Ewalds 1985). The obtained values normalized by $\sigma \sqrt{\pi R}$ are represented in Fig. \ref{Fig2}\,a as a function of the normalized crack length $a/R$ for different values of the ratio $w/R$. Both methods are found in agreement within $0.5\un{\%}$.

Following He et al. (1995) and Fett et al. (2005), the dimensionless stress intensity factor can be expressed as
\begin{equation}
{\sigma \sqrt{\pi R} \over K_I} = d_0 + d_1 {w \over R} + \left[d_2 {w \over R} + d_3 \right] {a \over R}.
\label{eq:EvansFett}
\end{equation}
The previous expression was used in He et al. (1995) to fit the simulations on the domain $2 \leq w/R \leq 4$ obtaining the parameter set $d_0 = 0$, $d_1 = 1$, $d_2 = 0.235$, and $d_3 = -0.259$. In Fett et al. (2005), the results were fitted on the same domain $2 \leq w/R \leq 4$, but taking also in consideration the effect of an offset $b$ of the hole with respect to the center of the sample ($b=0$ here) and obtaining $d_0 = -0.3703$, $d_1 = 1.1163$, $d_2 = 0.2160$, and $d_3 = -0.1575$. These two equations are plotted respectively in green and red in Fig. \ref{Fig2}\,b together with the results of our simulation. While the agreement is correct for some specific values of $w/R$ (close to 4 for the expression of He et al. (1995), closer to 3 for Fett et al. (2005)), some differences are observed, especially when reaching the extreme values of $w/R$.

This difference is attributed to the fact that Eq. (\ref{eq:EvansFett}) does not provide an accurate fit on the whole parameter range. However, an excellent fit can be obtained for a larger parameter range $2.5 \leq w/R \leq 5$ by allowing for a second order dependence in $w/R$ using the following expression
\begin{equation}
{\sigma \sqrt{\pi R} \over K_I} = \left[c_0 + c_1 {w \over R} + c_2 \left({w \over R}\right)^2 \right]
                                + \left[c_3 + c_4 {w \over R} + c_5 \left({w \over R}\right)^2 \right] {a \over R}
\label{eq:KInew}
\end{equation}
which leads in our dataset to the parameters $c_0 = 0.3156$, $c_1 = 0.7350$, $c_2 = 0.0346$, $c_3 = -0.4093$, $c_4 = 0.3794$, and $c_5 = -0.0257$. The agreement is excellent as shown by the comparison with the results of the simulations (blue curves) presented in Fig. \ref{Fig2}\,b.

The deviations from the fit for the largest values of $a/R$ are caused by the interaction of the crack with the specimen boundary and are beyond the domain of validity of the equation, which is given by $w <a <L-2w$. This regime with rapid increase of the stress intensity factor with crack extension corresponds to the sudden and instable crack propagation observed in the experiments prior to complete failure of the DCDC specimens in two symmetric pieces.

\subsection{Williams expansion series of the crack opening profile}

\begin{figure}[!h]
\begin{center}
\includegraphics[width=6cm]{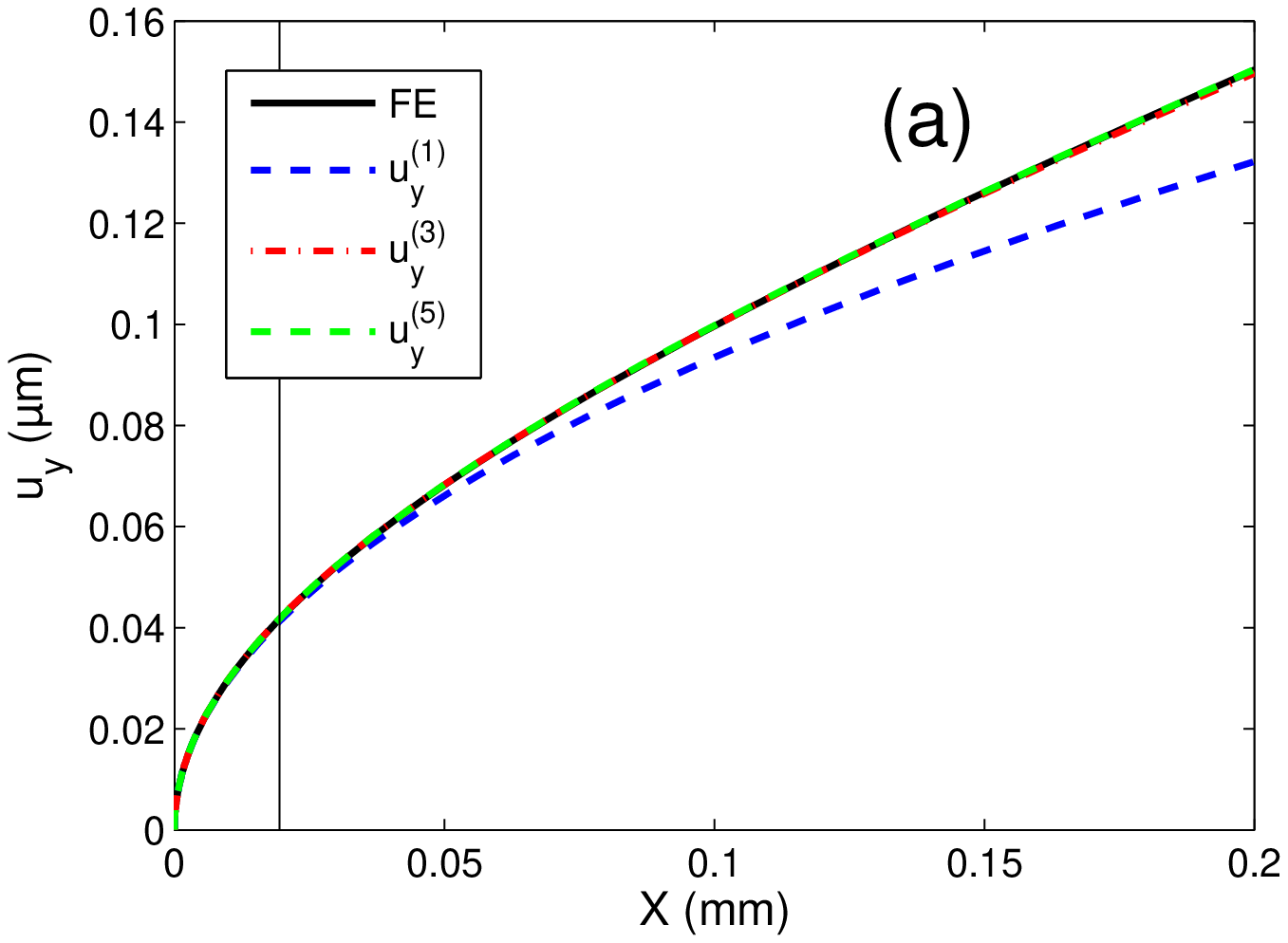}
\includegraphics[width=6cm]{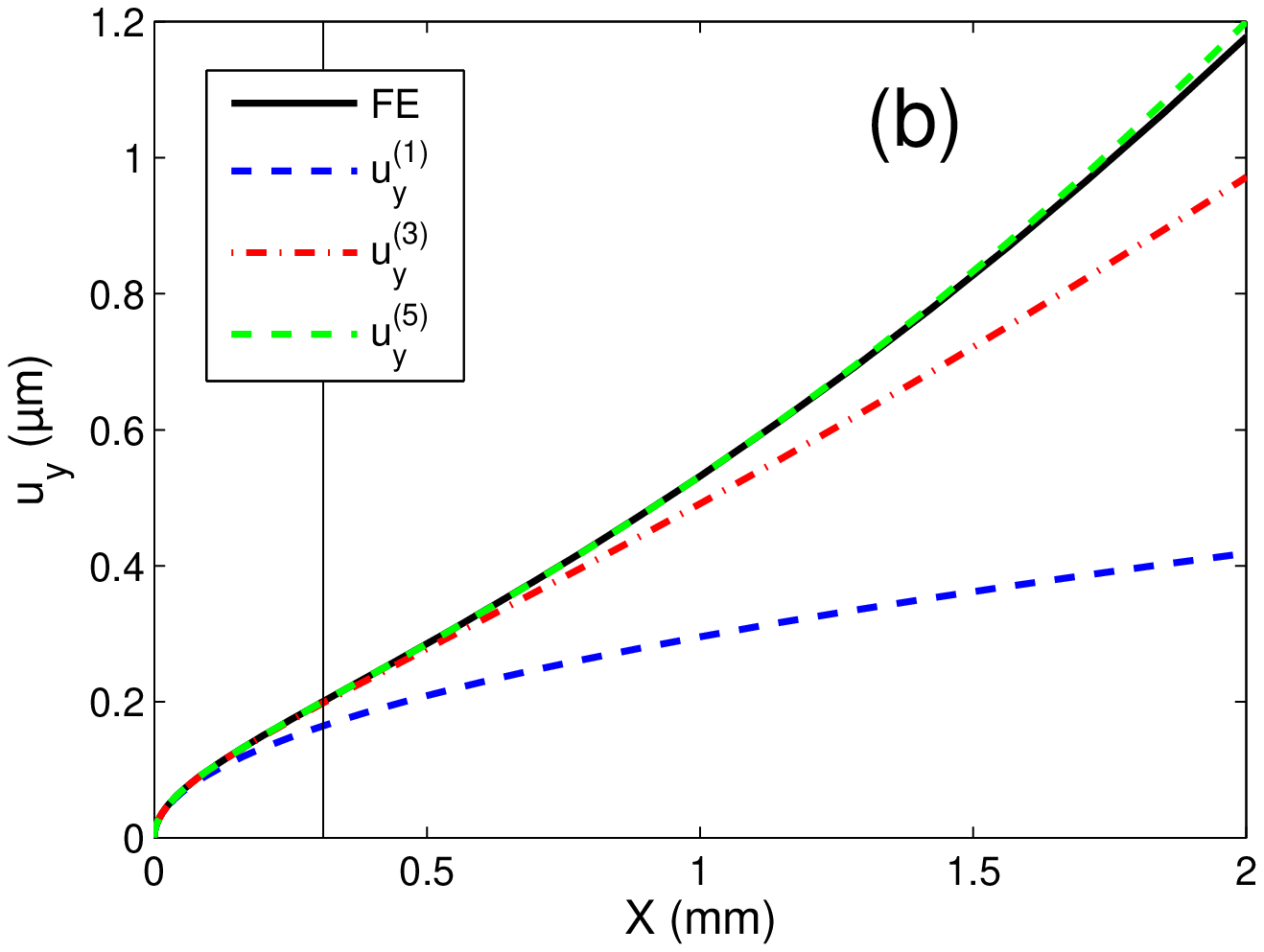}
\includegraphics[width=6cm]{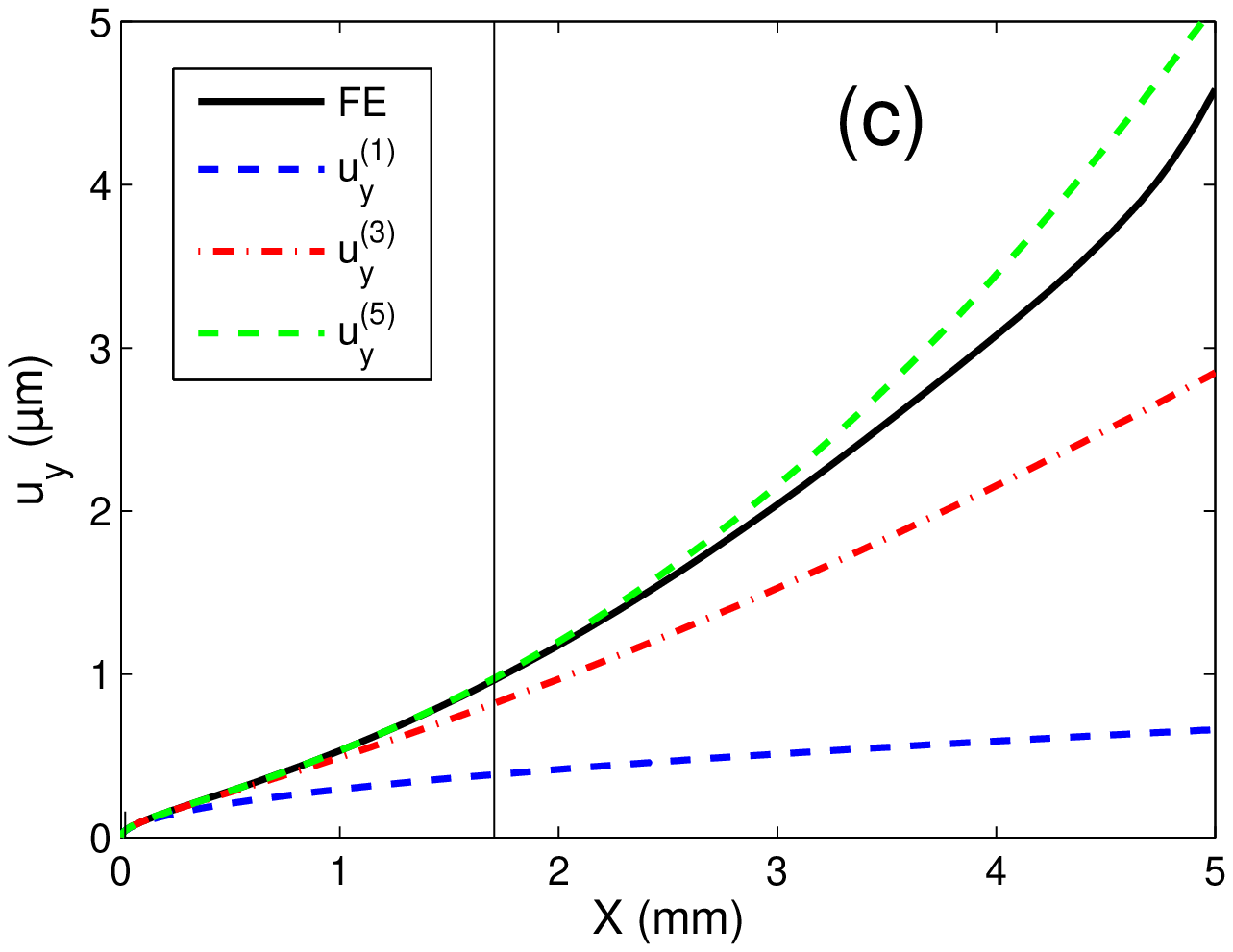}
\caption{Comparison between the crack opening profile obtained by FE simulation and the Williams series of order 1, 3, and 5 at different scales: (a) at the $\unm$ scale; (b) at the millimeter scale; (c) on the whole length of the crack. The vertical lines at $15\unm$, $300\unm$ and $1.7\un{mm}$ represent the $1\un{\%}$ limit of validity of $u_y^{(1)}$, $u_y^{(3)}$ and $u_y^{(5)}$, respectively}
\label{Fig3}
\end{center}
\end{figure}

The crack opening profile $u_y(X)$ is first analyzed in the same geometry as the samples used in the experimental part of this study. In particular, we choose $R = 0.531\un{mm}$, $w=2\un{mm}$ and $L = 20\un{mm}$, with an external applied stress $\sigma = 100\un{MPa}$. The crack opening profile obtained for a crack length $a = 5\un{mm}$ is shown in Fig. \ref{Fig3}. From the blowup (c), we can observe that the $1\un{\%}$ range of validity of the Irwin equation (blue dashed line) is limited to the first $20\unm$ near the crack tip. In order to provide an expression with a larger range of validity, $u_y(X)$ can be expressed in terms of a Williams expansion (Williams 1957; Maugis 1999)
\begin{equation}
u_y^{(n)}(X) = p_1 X^{1/2} + p_3 X^{3/2} + p_5 X^{5/2} + ... + p_n X^{n/2}
\label{eq:WilliamsOpening}
\end{equation}
where only the odd terms $p_1$, $p_3$, $p_5$,... $p_n$ are present due to the symmetry conditions of the pure mode I loading (the displacement function $u_y(x,y)$ must be antisymmetric in relation to the transformation $y \rightarrow -y$). The first term of this development corresponds to the Irwin expression given in Eq. (\ref{Eq1}). As a result, the first coefficient $p_1$ is directly associated with the singular stress behavior and can be expressed in terms of the stress intensity factor
\begin{equation}
p_1 = \frac{K_I}{E'}\sqrt{\frac{8}{\pi}}.
\label{eq:Williamsp1}
\end{equation}
As shown in the previous section, $K_I$ (resp. $p_1$) can be expressed as a function of the external stress and the geometrical parameters using the Eq. (\ref{eq:KInew}).

The other coefficients entering in the development of the crack profile are obtained by an iterative fit procedure: At first, the development to the first order $u_y^{(1)}$ is subtracted to the crack profile $u_y$ and is fitted by a function varying as $X^{3/2}$, providing the value of the coefficient $p_3$. The procedure is then iterated fitting $u_y-u_y^{(1)}-u_y^{(3)}$ by a function evolving as $X^{5/2}$, providing the value of the coefficient $p_5$. At this fifth order, the agreement with the results of the simulation is excellent up to $X \simeq 1.7\un{mm}$ as shown in Fig. \ref{Fig3}\,c. As a result, we limit our analysis to this order in the following of this study. The limitations of the developments of lower orders are presented in Fig. \ref{Fig3}\,a and \ref{Fig3} b. In particular, the Irwin equation \--- development to the first order \--- is found to be valid for $X < 15$ $\mu$m. The range of validity of these Williams developments is not universal and does depend on the sample dimensions as discussed in section \ref{sec:discussion}.

\begin{figure}[!h]
\begin{center}
\includegraphics[width=8.4cm]{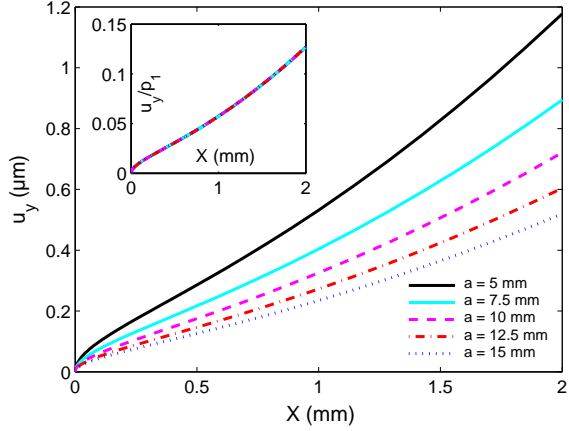}
\caption{Crack opening profiles obtained for different crack lengths. Once normalized by $p_1$, proportional to the stress intensity factor (see Eq. (\ref{eq:Williamsp1})), all the profiles follow the same master curve}
\label{Fig4}
\end{center}
\end{figure}

The effect of the crack length on the crack opening profile is now investigated. Fig. \ref{Fig4} represents the profile obtained for five values of $a$ ranging from $5\un{mm}$ to $15\un{mm}$. As shown in the inset of Fig. \ref{Fig4}, all profiles have the same shape once normalized by the first coefficient $p_1 = \frac{K_I}{E'}\sqrt{\frac{8}{\pi}}$ of their Williams development. In other words, the crack opening profiles can be written as
\begin{equation}
u_y(X) = \frac{K_I}{E'}\sqrt{\frac{8}{\pi}} \left( X^{1 / 2} + \alpha_3 X^{3 / 2} + \alpha_5 X^{5 / 2} \right)
\label{eq:WilliamsOpening2}
\end{equation}
where the coefficients $\alpha_3 = \frac{p_3}{p_1}$ and $\alpha_5=\frac{p_5}{p_1}$ are independant of the crack length.

\begin{figure}[!h]
\begin{center}
\includegraphics[width=8.4cm]{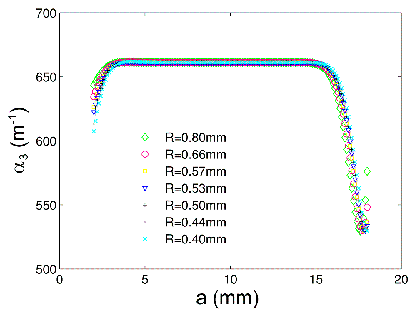}
\includegraphics[width=8.4cm]{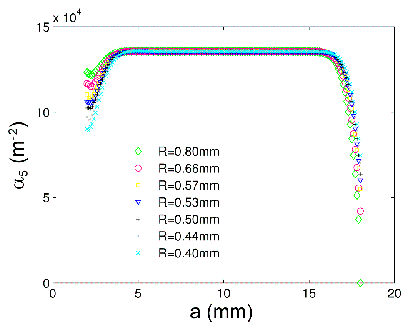}
\caption{Normalized coefficients $\alpha_3$ and $\alpha_5$ involved in the Williams development of the crack opening profile (see Eq. (\ref{eq:WilliamsOpening2})). If the crack tip is sufficiently far from the specimen sides, these coefficients, and therefore the shape of the opening profile, are independent of both the crack length $a$ and the radius $R$ of the hole of the DCDC specimens}
\label{Fig5}
\end{center}
\end{figure}

Very interestingly, these coefficients are also independent of the radius of the hole in the DCDC specimens. This is shown in Fig. \ref{Fig5} where the variations of $\alpha_3$ and $\alpha_5$ with the crack length are plotted for seven specimen geometries with different radius $R$, keeping constant the specimen width $w=2\un{mm}$. All the curves present an extended plateau regime as far as the crack tip is not too close to the specimen sides, as expected from the previous observations. Moreover, all the plateau regimes correspond to the same value of $\alpha_3$ and $\alpha_5$, irrespective of the radius of the hole. In other words, hole radius and crack length have no influence on the shape of the crack lips in a broad range of these geometrical parameters.

\begin{figure}[!h]
\begin{center}
\includegraphics[width=6cm]{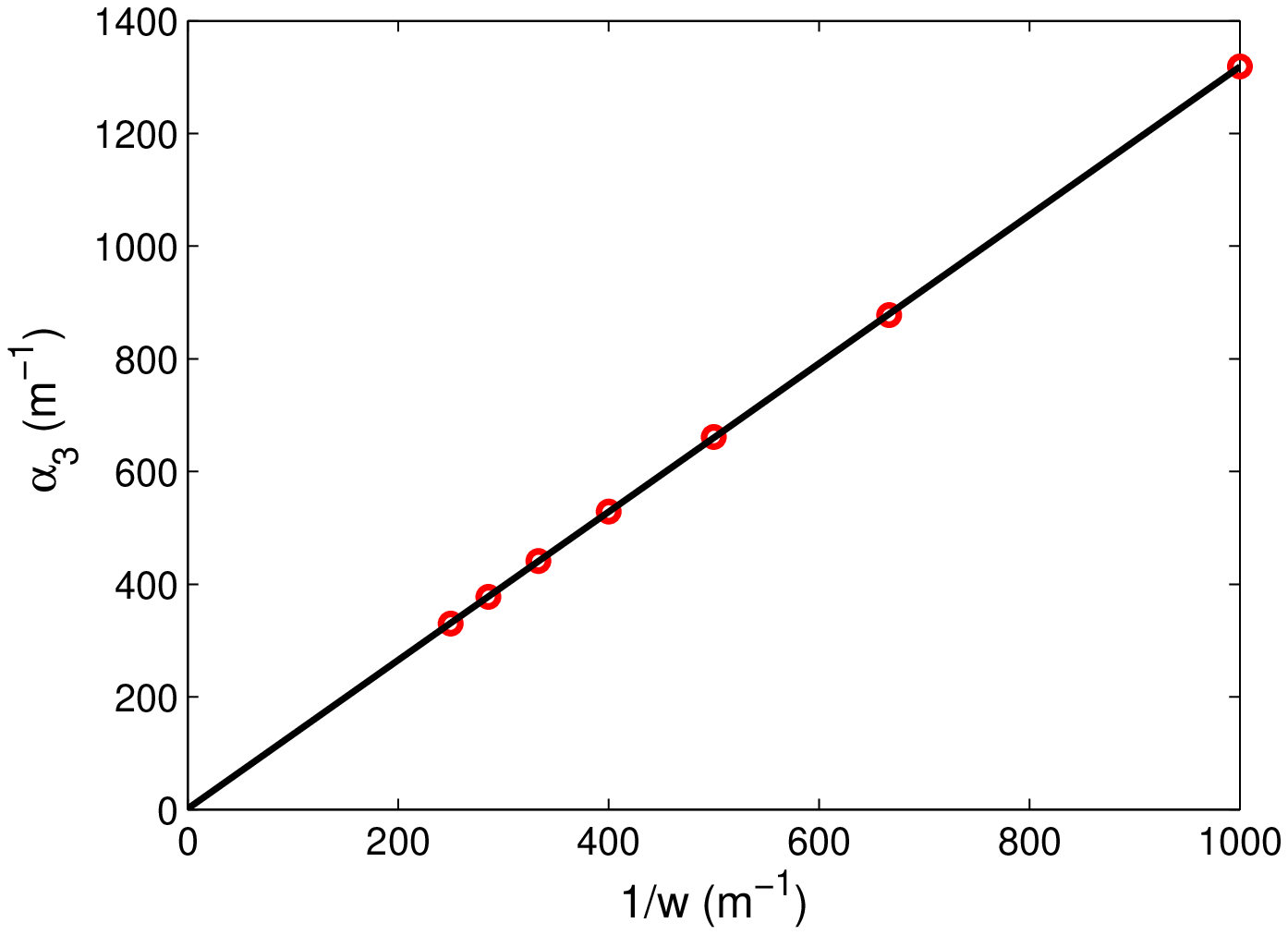}
\includegraphics[width=6cm]{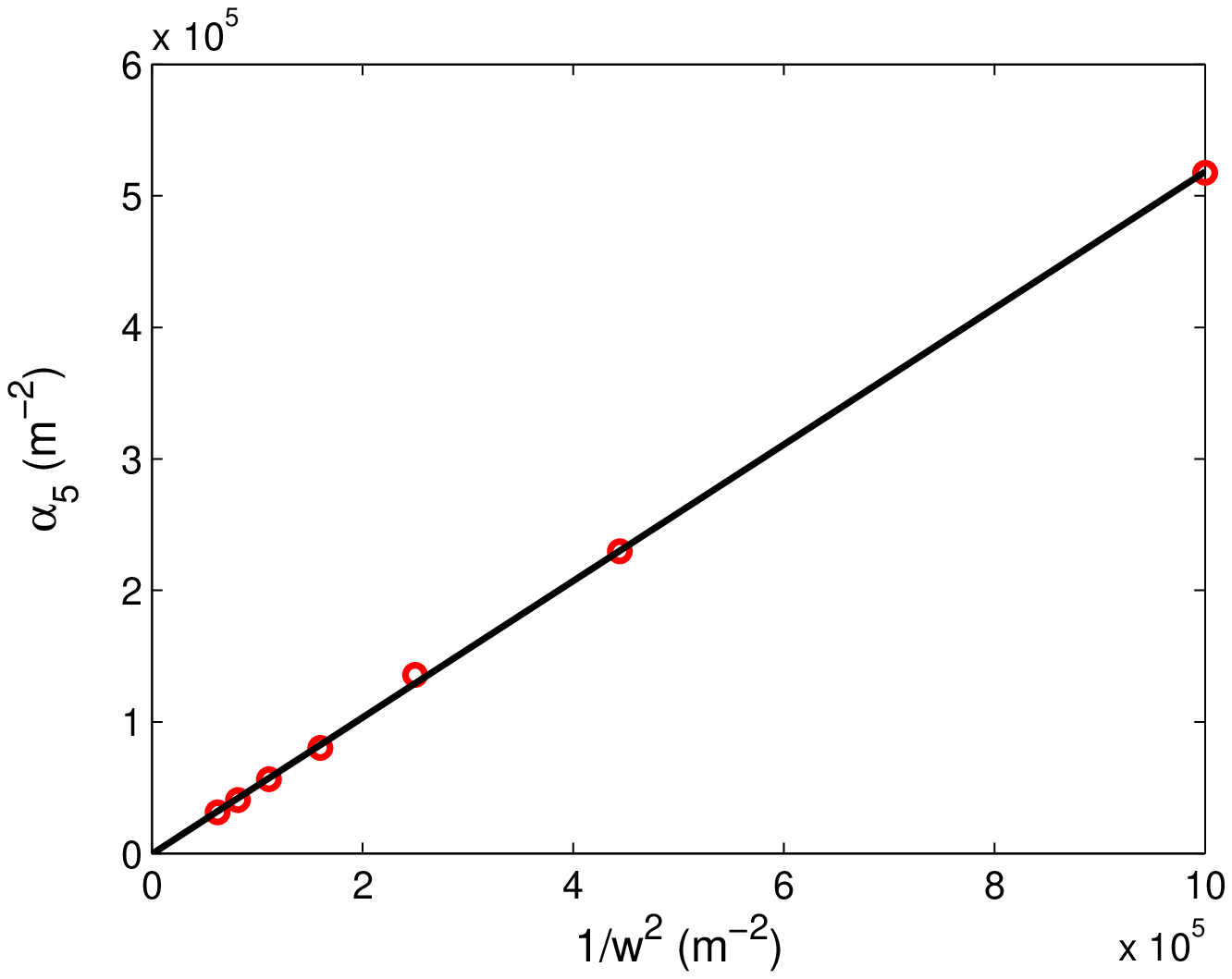}
\caption{Variations of the normalized coefficients $\alpha_3$ and $\alpha_5$ involved in the Williams development of the crack opening profile (see Eq. (\ref{eq:WilliamsOpening2})) with the specimen width $w$. The straight lines corresponds to variations as $\alpha_3 \sim 1/w$ and $\alpha_5 \sim 1/w^2$ as given in Eq. (\ref{Eq_alpha})}
\label{Fig6}
\end{center}
\end{figure}

However, a dimensional analysis suggests that both coefficients must vary with a length of the problem. More precisely, $\alpha_3$ and $\alpha_5$ are expected to scale as the inverse of a length and the inverse of the square of a length, respectively. The specimen width being the only remaining relevant length scale of the problem, it is natural to observe on Fig. \ref{Fig6} the following scalings
\begin{equation}
\alpha_3 = \frac{1.319}{w} \quad \mathrm{and} \quad \alpha_5 = \frac{0.515}{w^2}
\label{Eq_alpha}
\end{equation}
where the dimensionless constants are provided by linear regressions represented by straight lines.

As a result, combining Eq. (\ref{eq:WilliamsOpening2}) with Eq. (\ref{Eq_alpha}), one gets a rather simple description of the crack opening profile
\begin{equation}
u_y(X) = \frac{K_I}{E'}\sqrt{\frac{8}{\pi}} \sqrt{X} \left( 1 + 1.319 \left(\frac{X}{w}\right) + 0.515 \left(\frac{X}{w}\right)^2 \right).
\label{Eq_maitresse}
\end{equation}

This expression remains valid with a precision better than $1\un{\%}$ as far as the effects of the finite size of the specimen are not relevant, \ie for $2.5\,w - R < a < L - 2.3\,w - R$, and relatively close to the crack tip, \ie for $X < 0.85\,w$, because we limited the Williams development to the fifth order.
%We can now easily estimate the domain $X<\xi_1$ of 1\% validity of the first order (Irwin) term by letting $1.319(X=\xi_1/w)=1/100$, which leads to the scaling:
%\begin{equation}
%\xi_1 = {w \over 132}
%\label{eq:xi1}
%\end{equation}

Using the Eq. (\ref{eq:KInew}) providing the expression of the stress intensity factor as a function of the geometrical parameters, one obtains here a closed form of the opening profile as a function of simply measurable quantities of the problems. This can be very useful during DCDC experiments. Indeed, combining this analytical expression with the experimental measurement of the crack opening profile using \eg the interferometry technique described in the next section can provide an easy way to assess accurately the stress intensity factor that drives the crack growth in DCDC experiments.

But before discussing in more details these applications, we compare now the main results of this analysis with experimental measurements. We focus on one single DCDC geometry and compare the opening profile measured experimentally with the result of the finite element simulations.

\section{Experimental measurement by crack opening interferometry}

\begin{figure}[!h]
\begin{center}
\includegraphics[width=6cm]{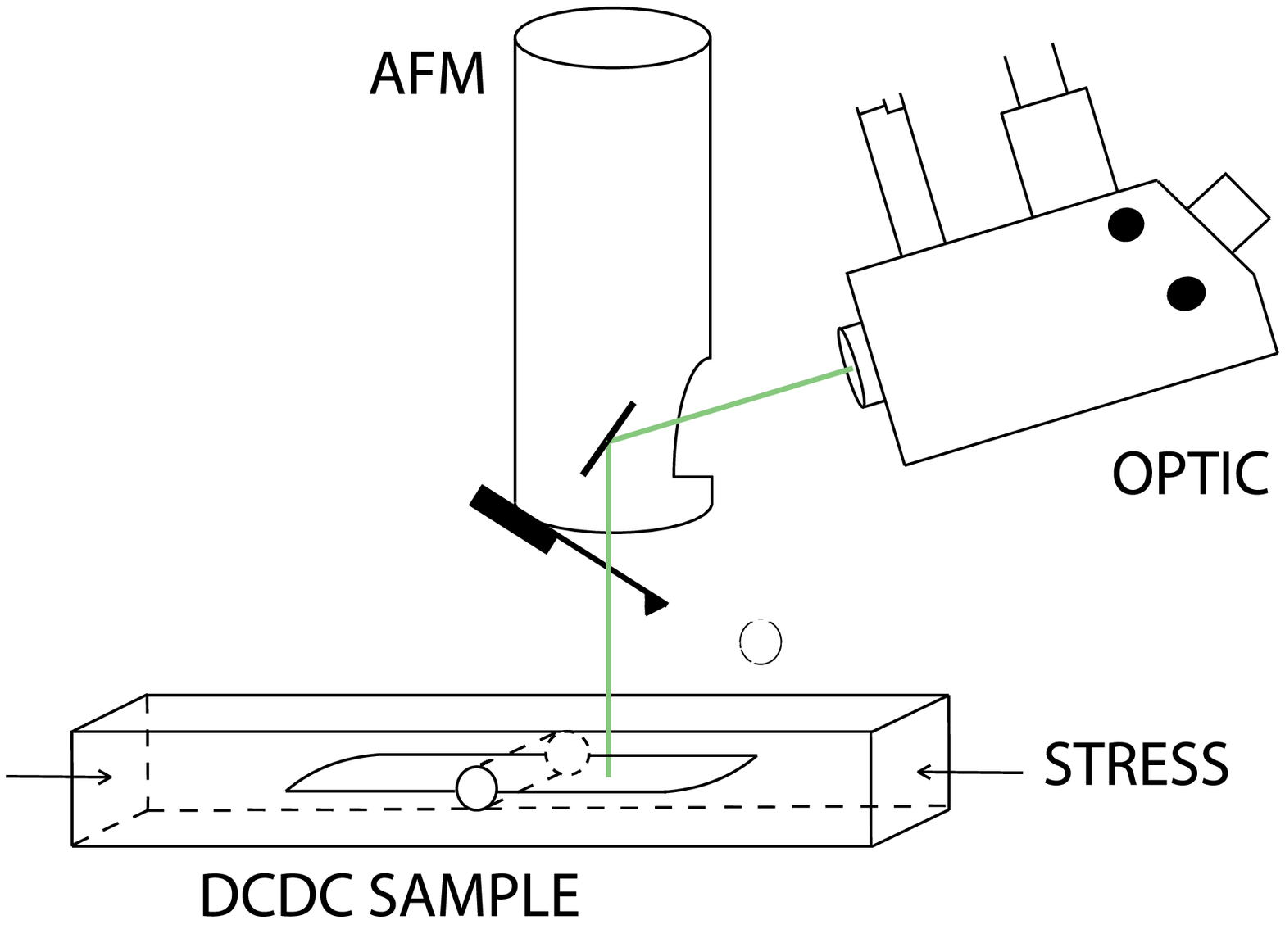}
\includegraphics[width=6cm]{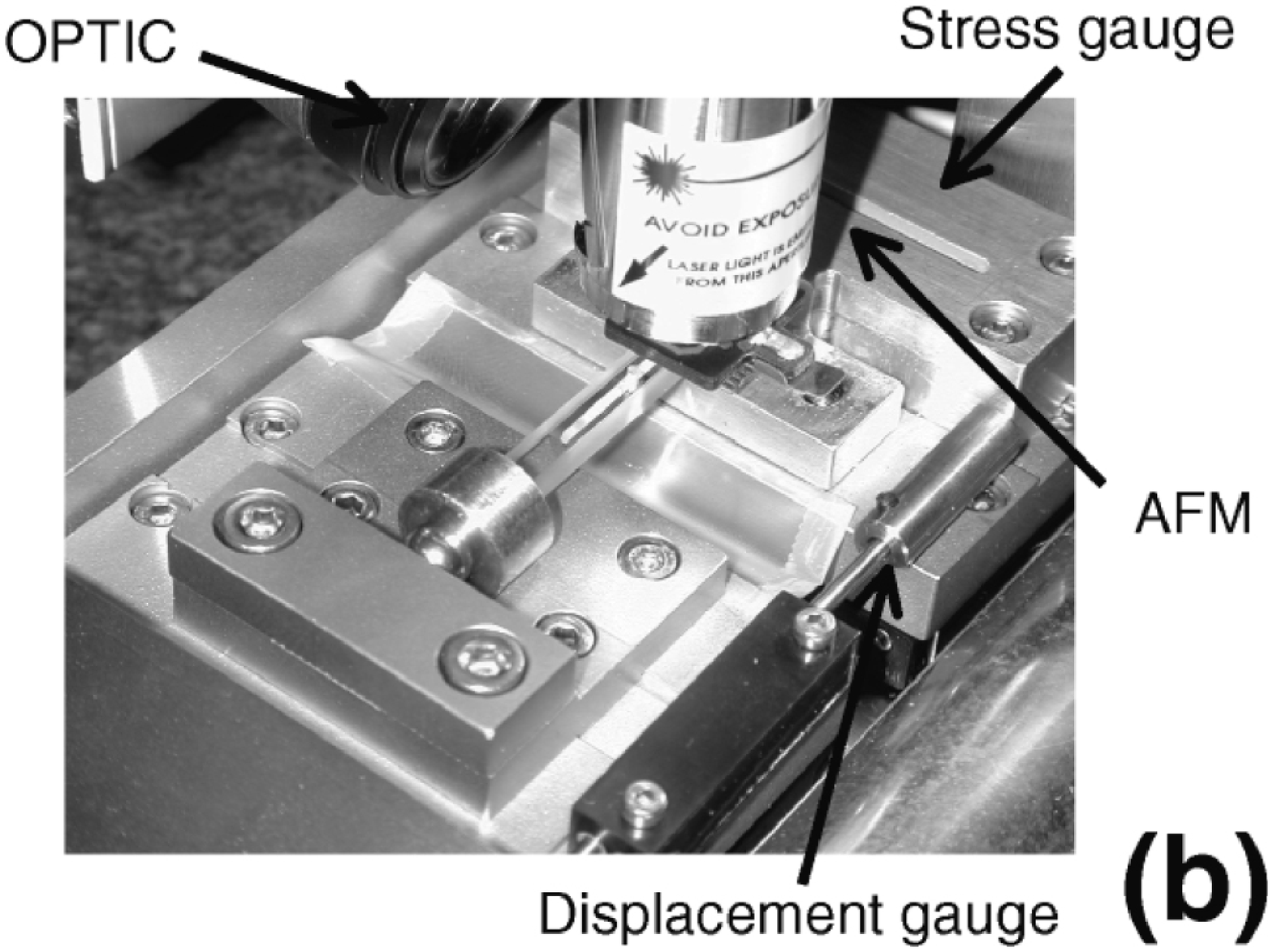}
\caption{Experimental setup: (a) Sketch of the DCDC sample arrangement in the crack opening interferometry technique (horizontal crack plane); (b) picture of the experiment in the AFM in-situ crack tip imaging mode (vertical crack plane)}
\label{fig:Fig7}
\end{center}
\end{figure}

The experimental measurement of the crack opening profile is performed by using a precision loading apparatus (based on a Microtest load cell produced by Deben, Woolpit, UK) providing highly stable failure conditions necessary for in-situ observations of slow crack propagations in glasses by Atomic Force Microscopy (AFM) (Célarié 2004). DCDC samples of pure silica glass (Suprasil 311, Heraeus, Germany) are machined into $4 \times 4 \times 40\un{mm^3}$ bars ($w=2\un{mm}$ and $L=20\un{mm}$) and polished with $CeO_2$ to a RMS roughness of $0.25\un{nm}$ for an area of $10 \times 10\unm^2$. A hole of radius $R = (0.531 \pm 0.010)\un{mm}$ was drilled at their center to trigger the start of two symmetric fractures (see Fig.\ \ref{fig:Fig7}).

\begin{figure}[!h]
\begin{center}
\includegraphics[width=6cm]{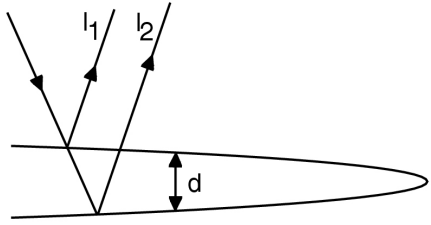}
\caption{Sketch of the crack opening interferometry technique. NB: exaggerated vertical scale for the crack opening and lateral scale for the reflected beams}
\label{fig:Fig8}
\end{center}
\end{figure}

The measurement of the crack opening profile is performed by means of monochromatic light reflection interferometry (Sommer 1970; Liechti 1993) according to the sketch of Fig. \ref{fig:Fig8}. When observing the crack tip by AFM at the free surface of the sample, the crack front is vertical as in Fig. \ref{fig:Fig7}\,b. In order to use the AFM setup to measure the crack opening displacement, the DCDC sample had to be placed with the fracture front lying horizontally. The white light source (illuminating vertically the sample after passing through the optical axis on the CCD camera and being reflected by an inclined mirror in the AFM head) was substituted by a green laser source (wavelength $\lambda=(532 \pm 1)\un{nm}$) in order to provide an intense monochromatic light (Fig.\ \ref{fig:Fig7} a).

When the monochromatic beam reaches the fracture plane in an
orthogonal direction, it is split into two beams that are
reflected by the two fracture surfaces and then collected back by
the CCD camera. The path difference between the two beams is twice
the distance between the two lips of the crack, \ie $\Delta = 2\,d
= 4\,u_y(X)$ according to the notations used previously. The phase
difference $\delta$ is therefore given by
\begin{equation}
\delta = \pi + \frac{2\pi n}{\lambda}4\,u_y(X)
\label{eq:PhaseDelay}
\end{equation}
where $n$ is the refractive index of the fluid filling the gap in
the fracture and the term $\pi$ is a constant additive phase shift
caused by the fact that both beams are normally reflected on
symmetrically opposite interfaces (respectively glass/fluid and
fluid/glass) (Born and Wolf 1999).

The intensity $I$ of the reflected light is given by:
%related to the local crack opening $d = 2u_y(X)$ by the relation:

\begin{equation}\label{eq:Intensity}
I = I_1 + I_2 + 2(I_1I_2)^{1/2} \cos(\delta) \simeq 4I_1 \cos^{2} \left( \frac{\delta}{2} \right)
\end{equation}
where the approximation on the right stems from the fact that the
two reflected beams have similar intensity. Since $u_y(X)$ is a
monotonically increasing function of $X$, the reflected intensity
will develop into a series of fringes parallel to the crack front,
each new one corresponding to an increase of the crack opening $d
= 2\,u_y(X)$ by a quantity $\lambda/2$. The term $\pi$ in Eq.
(\ref{eq:PhaseDelay}) implies the presence of an intensity minimum
at $X=0$. The first intensity maximum is obtained for $2\,u_y(X_1)
=\lambda/4$ and the position $X_k$ of the $k^{th}$ order fringe is
given by the equation

\begin{equation}\label{eq:FringePosition}
2 u_y(X_k) = \left( \frac{1}{2} + k \right) \frac{\lambda}{2}.
\end{equation}

\begin{figure}[!h]
\centering
\includegraphics[width=8.4 cm]{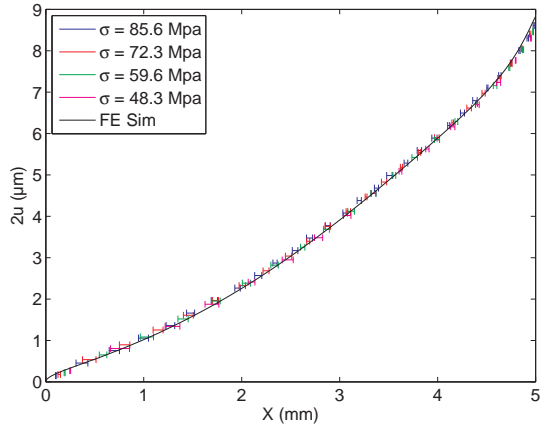}
\caption{Comparison between the opening crack profile obtained by the four experiments (in color, with errorbars) at different loads (renormalized) and the results of the FE simulation for the same set of parameters}
\label{fig:Fig9}
\end{figure}

By measuring the position and order of the fringes along the whole crack, we obtain a series of discrete data points on the full
crack opening profile as shown in Fig. \ref{fig:Fig9}.

In order to provide a better resolution, the fringe pattern along the $5\un{mm}$ crack length is imaged with a series of $1\un{mm}$
overlapping images. To insure that we preserve the right order and position of the fringes, the displacements of the load cell are
measured by a heterodyne interferometer (Zygo) with $10\un{nm}$ resolution. The uncertainty in the position of each fringe ($\sim 50$ $\mu$m) was estimated depending on the quality of the local signal and fringe shape (the error bars are plotted in Fig. \ref{fig:Fig9}). A special attention must be paid to the first fringe, since the opening gradient is maximal in that region. However, in our operating conditions, the position of the first
fringe is located to some hundreds of micrometers from the crack tip, making its measurement safe. The eventual optical bias induced by the
stress gradient on the apparent crack tip position can be shown to be less than 0.1 $\mu$m in glass samples (Kysar 1998).

In order to verify the robustness of the method and the linearity of the loading configuration and sample, the measurement was
repeated for four different loads ($\sigma = 50$, $62.5$, $75$ and $87.5\un{MPa}$) and the opening profile was normalized to the
reference load $\sigma = 100\un{MPa}$ also used in the FE simulations according to linear elasticity. The four experimental
series were realized in a time span of one hour in order to guarantee stable conditions ($T = (21 \pm 1)^\circ \un{C}$, $RH =
(35 \pm 5)\un{\%}$) and minimal crack propagation.
Since the highest stress used in this experiment corresponds to $K_I = 0.36\un{MPa\,m^{1/2}}$,
the maximum crack propagation velocity is of the order of $10^{-10}\un{m\,s^{-1}}$
(as determined by AFM in-situ crack propagation experiments effectuated on the same sample as shown in Fig. \ref{fig:Fig7} b, cf.\ Célarié 2004).
The maximum crack advance in one hour can thus be estimated to $0.4\unm$, which is below the resolution of our optical
microscope. The crack length can thus be considered as constant with the value $a = (5.02 \pm 0.01)\un{mm}$.

The results of Fig. \ref{fig:Fig9} show an excellent agreement
between the four measurements within the experimental errors. The
agreement is also excellent when comparing the experimental
measurements to the results of the FE simulation for the same set
of parameters (represented by a black line on Fig.
\ref{fig:Fig9}). We point out that the agreement with the FE
simulations is also excellent in the diverging part near the
central hole (corresponding to distances $X \simeq a = 5\un{mm}$
from the crack tip). We should remind that the FE simulation is
made in 2D under plane strain condition in order to represent the
deformation field in the bulk of the DCDC specimen where the
crack opening interferometry is applied. When we refer to the crack
length $a$ in a 3D specimen, this should thus be measured in the
center of the specimen.

Finally, due to the robustness of the crack opening profile shown in the previous section, the agreement between the FE simulation and the experimental measurements for a given sample geometry can be considered as representative for a broad set of geometrical parameters.

\section{Discussion}
\label{sec:discussion}

The expression (\ref{Eq_maitresse}) for the crack opening profile in the DCDC specimen shown in the theoretical part of this paper
and validated in the experimental section is quite interesting. The first important remark is that the fifth order Williams
expansion is invariant with respect to variations in the crack length and the hole radius. Once the stress intensity factor is
determined by Eq. (\ref{eq:KInew}), this acts simply as a multiplicative factor to a constant crack opening profile. Of
course, this is only valid for sufficiently long crack lengths ($a>2.5\,w-R$) and crack tips far enough from the specimen side
($a<L-2.3\,w-R$). In addition, this expression gives an accurate description of the opening profile in the domain of validity of
the fifth order Williams expansion $X < 0.85\,w$.

As a consequence, the ranges of validity of the first and third order developments are also invariant for samples of same width
$w$, and are expected to scale with $w$.
%\textbf{In fact, the caracteristical length of the system is given by $w$ shown in the expression (\ref{Eq_maitresse})}.
%By considering the ratio of the different terms in eq.\ (\ref{Eq_maitresse}), it can be easily found that
The range of $1\un{\%}$ validity of the first order term (Irwin equation (\ref{Eq1})) is given by $X < 0.0076\,w$, while the range of validity of the third order development is $X < 0.15\,w$.
An estimate of these upper bounds can be easily found comparing the terms of the expansion of the crack opening profile, as given
in Eq. (\ref{Eq_maitresse}).
%The constants involved in this expression being of the order of the unity, the third order term $\frac{X}{w}$ is comparable to $1\un{\%}$ of the first order term equal to unity for $X \simeq 0.01\,w$ while the fifth order term $\left(\frac{X}{w}\right)^2$ is as far as relevant for $X \simeq 0.1\,w$, following the same argument.

We should point out that the use of typical optical systems available in AFM instruments
does not allow the interferometric measurement of the close neighborhood of the crack tip.
The custom coupling of this system with a high quality optical and illumination system
would allow using higher magnifications and extracting valuable information from
the interpolation of the fringe intensity profiles (cf.\ Swadener and Liechti 1998).
However, the present results allow us to measure the stress intensity factor
by a simple and direct experimental measurement of the crack
opening displacement over the first millimeters from the crack tip.
In fact, fitting the experimental profile by the Eq.
(\ref{Eq_maitresse}) using two free parameters (position of the
crack tip and multiplicative constant of the profile), this
technique enables an accurate measurement of the stress intensity
factor and the crack length. This single measurement might be an
interesting alternative to the load measurement combined with the
optical estimation of the crack length necessary to assess
the stress intensity factor, as classically done for DCDC fracture
tests. This feature is particularly interesting for in-situ
investigations of the crack tip since it allows reducing the
weight and dimensions of the loading fixture.

We remind that, as discussed by Plaisted et al. (2006), linear modelling can not provide a complete description of the DCDC deformation field, since the second order terms due to rotation can become very relevant, especially for long cracks. However, these effects are relatively limited in very stiff and brittle materials and do not seem to affect the present analysis as confirmed by the excellent agreement between the linear analysis and the experiments realized at four different loads. Moreover, subcritical propagation $K(v)$ curves performed repeatedly on different crack length regions of the same DCDC glass specimen are found to be in satisfactory agreement between them (Célarié 2004), thus confirming the validity of the present treatment for glasses. A generalization of our approach to softer materials by accounting for nonlinear terms would be an interesting future development.

Finally, it is interesting to discuss the limitations of our
approach at much smaller scales. Due to the elevated degree of
homogeneity and brittleness of silica glass, the lower limit of
validity of the Irwin relation for a sharp crack is expected to be
of the order of a few nanometers from the crack tip. In addition,
the FE simulation is strictly a good representation of a
measurement of the crack opening in vacuum. In ambient condition,
capillary condensation is expected to develop into the crack
cavity (Wondraczek et al. 2006; Grimaldi et al. 2008; Ciccotti et
al. 2008). On the one hand, this can bias the apparent position of
the crack front due to the weak reflectivity of the air/glass
interface. On the other hand, this can alter locally the crack
opening profile due to the attractive capillary forces. For the
experiments presented here, the largest length of the capillary
condensed phase can be estimated to few micrometers. As a
consequence, this corresponds to a distortion of the crack lip on
the first few micrometers close to the tip, which can
unfortunately not be measured by the present setup, but which could
be revealed by the coupling it with a higher quality optical and illumination system.
The effect of capillary condensed phase represents an interesting challenge
for future experimental and theoretical investigations.

\section{Conclusion}

This work reports a combined theoretical and experimental study of
the crack opening profile in DCDC specimens of linear elastic brittle
materials. It has been shown how a fifth order Williams expansion
series can be used to describe very accurately the crack opening
profile over a conveniently large domain near the crack tip.
Finite element simulations have allowed on one side to obtain an
accurate expression of the variations of the stress intensity
factor with the crack length for a broad range of DCDC geometries,
and on the other side, to express the crack opening profile in a
simple form as a function of the geometrical parameters of the
specimen. The ranges of validity of these analytical expressions
and more specifically of the Williams development at various
orders are calculated and interpreted. Moreover, the simulated
profiles were successfully tested against an accurate experimental
measurement by monochromatic-light crack opening interferometry.
This single optical measurement might be an
interesting alternative to the load measurement combined with the
optical estimation of the crack length both necessary to assess
the stress intensity factor, as classically done for DCDC fracture
tests.

\begin{acknowledgements}
We wish to thank C. Marlière and F. Célarié for initial developments of the fracture experiment and C. Blanc for his help in the setup of the interferometric technique. We thank S. Roux and T. Fett for stimulating discussions. This work was supported by the ANR grant CORCOSIL (BLAN07-3-196000). \\
\end{acknowledgements}

\end{document}